\documentclass[prb,amssymb, amsmath, showpacs,superscriptaddress]{revtex4-1}
\usepackage{graphicx}
\usepackage{dcolumn}
\usepackage{bm}
\usepackage[colorlinks=true,linkcolor=blue]{hyperref}
\begin{document}

\title{Diffusion of Mn interstitials in (Ga,Mn)As epitaxial layers}

\author{L. Hor\'{a}k}
\email{horak@karlov.mff.cuni.cz}
\author{J. Mat\v{e}jov\'{a}}
\author{X. Mart\'{i}}
\author{V. Hol\'{y}}
\affiliation{
Department of Condensed Matter Physics, Charles University,
Prague, Czech Republic}

\author{V. Nov\'{a}k}
\affiliation{ Institute of Physics ASCR, Prague, Czech Republic }
\author{Z. \v{S}ob\'{a}\v{n}}
\affiliation{
Department of Microelectronics,
The Czech Technical University in Prague, Czech Republic}
\affiliation{ Institute of Physics ASCR, Prague, Czech Republic }

\author{S. Mangold}
\affiliation{Karlsruhe Institute of Technology, Karlsruhe,
Germany}

\author{F. Jim\'{e}nez-Villacorta}
\affiliation{Department of Chemical Engineering, Northeastern University,
Boston, USA}

\date{\today}

\begin{abstract}
Magnetic properties of thin (Ga,Mn)As layers improve during
annealing by out-diffusion of interstitial Mn ions to a free
surface. Out-diffused Mn atoms participate in the growth of a
Mn-rich surface layer and a saturation of this layer causes an
inhibition of the out-diffusion. We combine high-resolution x-ray
diffraction with x-ray absorption spectroscopy and a numerical
solution of the diffusion problem for the study of the
out-diffusion of Mn interstitials during a sequence of annealing
steps. Our data demonstrate that the out-diffusion of the
interstitials is substantially affected by the internal electric
field caused by an inhomogeneous distribution of charges in the
(Ga,Mn)As layer.
\end{abstract}


\maketitle

\section{\label{introduction}Introduction}

The coupling of transport and magnetic properties in (Ga,Mn)As
thin films has triggered an intense research in the field of
spintronics.\cite{Science-2000-Dietl-1019-22} Most of the
efforts have been focused to increase the Curie temperature
($T_{\mathrm{C}}$) from the approx. 40 K in as-grown films up to
the current record temperature of approx. 190 K, being the most
successful strategy the post-growth
annealing.\cite{PhysRevLett.92.037201, PhysRevB.78.054403,
PhysRevB.78.075319} In spite of the numerous theoretical and
experimental works, a successful modeling of the involved
processes is still lacking.

In an ideal approach, Ga atoms (approximately up to 10\%) of the
host GaAs lattice are substituted by Mn atoms
(Mn$_{\mathrm{sub}}$) and supply both charge (act as acceptors)
and magnetic moments that order ferromagnetically. Far from this
scenario, some of the Mn atoms are located in interstitial
positions (Mn$_{\mathrm{int}}$)\cite{PhysRevB.65.201303} and
supply two electrons and order antiferromagnetically to
Mn$_{\mathrm{sub}}$.\cite{PhysRevB.71.064418} As a result of the
competing processes, $T_\mathrm{C}$ and saturation magnetization
decrease.\cite{PhysRevB.65.235209,RevModPhys.78.809} The
reduction of point defects cannot be achieved by increasing the
growth temperature (like in the case of GaAs thin layers) due to
the formation of metallic MnAs
clusters.\cite{PhysRevLett.63.1849} Instead, the (Ga,Mn)As
layers are post-growth annealed to decrease the density of
interstitials\cite{PhysRevLett.92.037201,PhysRevB.65.201303} and,
 to achieve the current record $T_\mathrm{C}$, the annealing is
combined with successive surface
etching.\cite{PhysRevB.78.054403} Recent studies indicate that
Mn$_{\mathrm{int}}$ diffuse through the (Ga,Mn)As lattice
towards the free surface where they are passivated and produce a
thin Mn rich surface layer.\cite{PhysRevB.78.054403,
PhysRevB.78.054403} However, the lack of non-destructive probing
methods has prevented a step-by-step monitoring of the process
across the whole layer and its detailed modeling.

In our previous paper we described a method for the
determination of the concentrations of Mn ions in various
lattice positions, the method is based on a precise measurement
of high-resolution x-ray diffraction (HRXRD) curves around
several reciprocal-lattice points.\cite{predchozi-clanek} A
variant of the HRXRD method consists in the measurement of the
dependence of the diffracted intensity on the photon energy
around the MnK absorption edge (anomalous diffraction). From
this dependence it is possible to determine the difference of
the densities of Mn atoms placed in non-equivalent interstitial
lattice positions.\cite{HolyAPL2010}

Another x-ray based methods for the investigation of the lattice
positions of Mn atoms in (Ga,Mn)As are the x-ray standing wave
method (XRSW) and x-ray absorption spectroscopy (methods
extended x-ray absorption fine structure -- EXAFS and x-ray
absorption near-edge spectroscopy -- XANES). The former method
uses the effect of x-ray standing wave produced by diffraction
in the GaAs substrate. The standing wave excites a fluorescence
of the Mn atoms and from the dependence of the fluorescence
intensity on the incident angle of the primary radiation the Mn
lattice positions can be deduced.\cite{PhysRevB.74.245205,
LeePRB2010} From the XRSW data it follows that the density of Mn
interstitial atoms indeed decreases during annealing, however a
quantitative determination of the Mn density profile is
practically not possible. The EXAFS method makes it possible to
determine roughly the relative amount of the Mn interstitials in
a (Ga,Mn)As layer (see Refs. \onlinecite{BacewiczJPCS2005,
0953-8984-19-49-496205}, among others).

Here we present a detailed numerical model of the process of
out-diffusion of Mn$_{\mathrm{int}}$ during annealing supported
by HRXRD and X-ray absorption near edge spectroscopy (XANES). We
investigate a sample of (Ga,Mn)As after several annealing steps
combined with a wet-chemical removal of the surface layer. The
depth profiles of the concentration of Mn$_{\mathrm{int}}$
obtained from the HRXRD measurements are compared with the
results of numerical simulation of Mn diffusion. From this
comparison we are able to estimate the diffusion coefficient of
the interstitial ions as well as their trapping rate at the
surface. In addition, we study the Mn-rich layer surface by
XANES using a grazing incidence surface-sensitive geometry. From
the experiments it follows that the surface layer acts as a sink
for the incoming Mn interstitials and the out-diffusion process
is substantially affected by local electric field in the (Ga,Mn)
As layer.

\section{\label{Experiment}Experiment}
The (Ga,Mn)As epitaxial layer of the nominal thickness of 100~nm
and the Mn content of 14\% has been grown on a GaAs buffer with
the thickness of 150~nm deposited on a GaAs(001) substrate.
After the deposition the sample was cleaned in O-plasma and HCl
and cleaved to two parts. The first part was "as-grown" control
sample (A) with the Curie temperature below 6~K. The second part
was annealed in the air for 24 hours at 160$^\circ$C, the Curie
temperature increased to 142~K (sample denoted as B-1).

Both samples were characterized by HRXRD. According to the
method described in our previous paper\cite{predchozi-clanek} we
measured diffraction curves around several reciprocal lattice
points. The results indicated (see the following sections) that
not all of the interstitial Mn atoms diffused out during the
annealing, most likely the passivation of the surface by
interstitial ions inhibited the diffusion process. To make
evident that the remaining Mn interstitials in the (Ga,Mn)As
layer are still mobile, the sample B-1 was rinsed in HCl for
30~s to remove the surface Mn-rich layer and subsequently
annealed at 160$^\circ$C for one hour. This cleaning\&annealing
procedure has been repeated 20 times. The sample after the
twenty steps of the cleaning\&annealing procedure (denoted as
B-2) was finally annealed at 160$^\circ$C for 80 hours to
homogenize the layer and the sample in the final state was
denoted as B-3. The sample in states B-2 and B-3 were
characterized again by the HRXRD procedure described in Ref.
\onlinecite{predchozi-clanek} to obtain the depth profile of the
concentration of the Mn interstitials. The experimental
diffraction curves for all samples and all diffractions were
simultaneously fitted with theoretical curves. The same value of
the concentration of the substitutional Mn was kept for all
samples during the fitting procedure, since we assume that the
Mn atoms in substitutional positions are not affected by the
annealing.

The HRXRD measurements were carried out by a standard laboratory
high-resolution x-ray diffractometer equipped with a 2~kW Cu
tube with linear focus, parabolic multilayer mirror and a
4$\times$Ge220 Bartels monochromator. We used a three-bounce Ge
analyzer crystal in front of a point detector. Examples of the
measured and fitted diffraction curves are presented in Fig.
\ref{detail}, the fitting procedure and the results are
commented in Section \ref{HRXRDResults}.

The XANES measurements have been carried out at the ANKA
synchrotron source, Karlsruhe. X-Spec ionisation chambers from
Oxford-FMB were used for the intensity monitors and a five
element germanium detector from Canberra for the detection of
the fluorescence energy. The used peaking time of the detector
electronic was 500 ns. We measured the dependence of the
intensity of the MnK$\alpha$ fluorescence line in dependence of
the energy of the primary x-ray beam for various angles
$\alpha_\mathrm{i}$ below and above the critical angle
$\alpha_\mathrm{c}$ of total external reflection,
$\alpha_\mathrm{c} \approx 0.38$~deg for the energy range used
(around 6.5~keV). The surface sensitivity of the XANES
measurement is achieved by the fact that increasing
$\alpha_\mathrm{i}$, the penetration depth and consequently the
depth from the fluorescence signal is collected, increases from
few nm for $\alpha_\mathrm{i}<\alpha_\mathrm{c}$ to several
$\mu$m.

\section{\label{HRXRDResults}Results of the diffraction measurements}

The experimental HRXRD curves of all samples were fitted using a
standard two-beam dynamical theory of x-ray diffraction
\cite{holy-bible}. From the fit we obtained the optimized
parameters (the concentration of the interstitials and the
thickness of the layer), their values are summarized in Tab.
\ref{tabulka}.
The details of the experimental data evaluation
are described in the previous paper\cite{predchozi-clanek}.

The substitutional Mn atoms are not affected by the annealing
\cite{PhysRevB.65.201303}, this allows us to consider the same
concentration of the substitutional Mn in all samples during the
fitting procedure. The concentration of the substitutional Mn
($c_\mathrm{sub} = (8.2\pm 1.1)\%$) was determined by anomalous
x-ray diffraction method performed on the same
samples.\cite{HolyAPL2010}

\begin{figure} 
\includegraphics[keepaspectratio,width=8cm,clip]{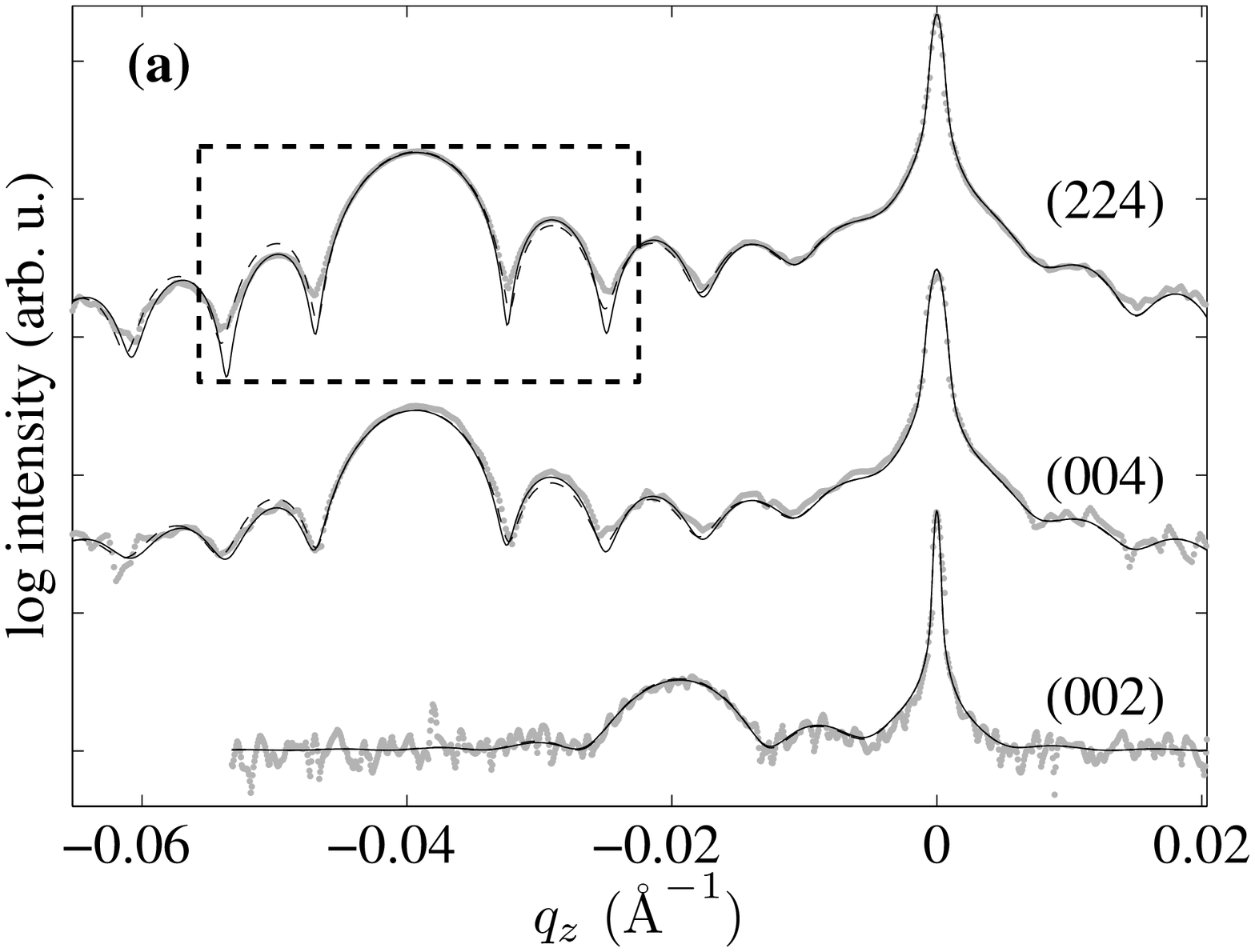}
\includegraphics[keepaspectratio,width=8cm,clip]{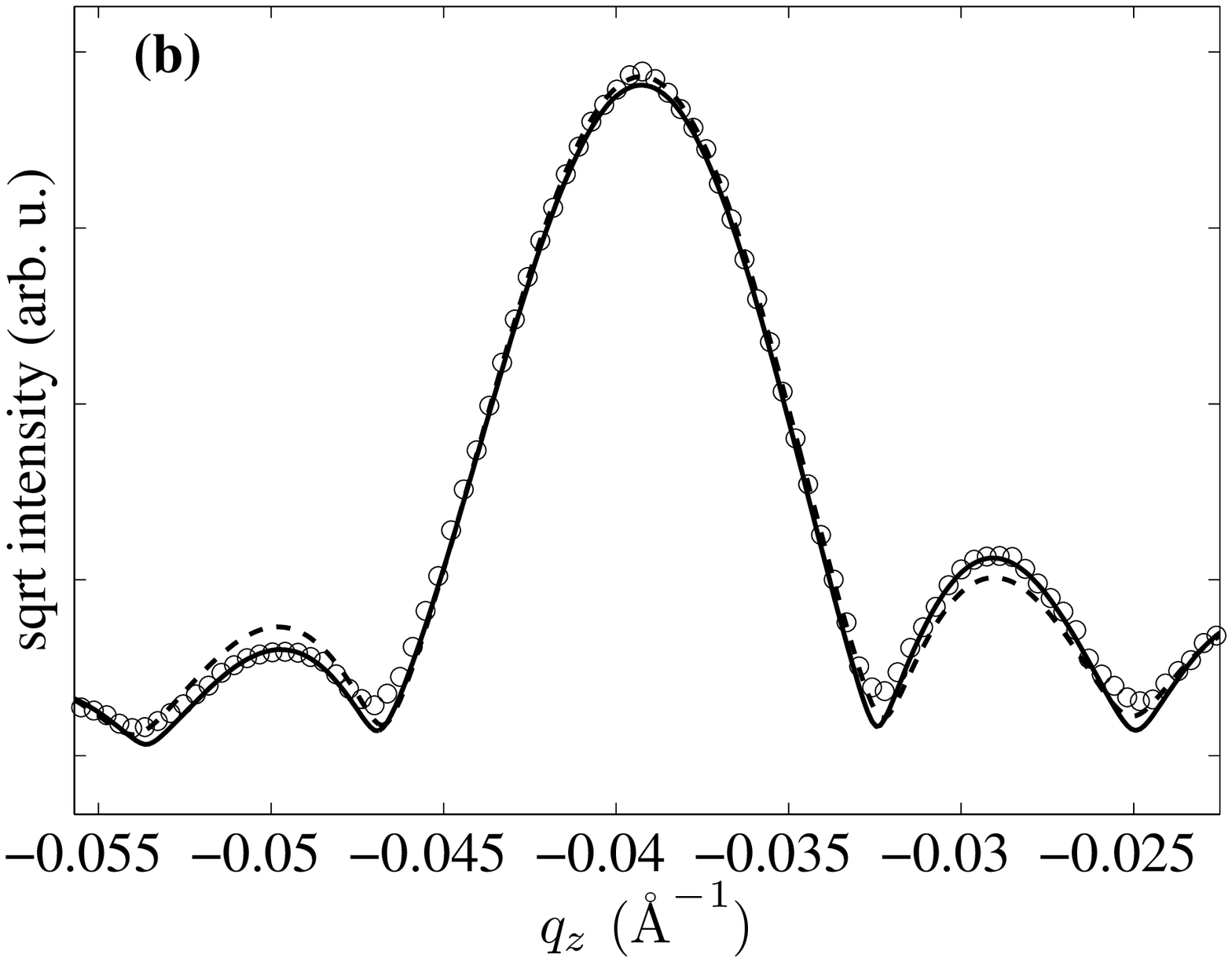}
\caption{\label{detail} (a) Diffraction curves of sample B-2 (gray
points) and fitted simulated curves using a single-layer model
(dashed line) and a bilayer model (full line). The diffracted
intensity is plotted as a function of the vertical component $Q_z$
of the scattering vector. The part of the figure marked by the
dashed box is enlarged in panel (b), where the diffraction maximum
of the layer is plotted in detail; here the measured data are
represented by circles. The diffraction curve simulated using the
bilayer model fits better the experimental data than the
single-layer model, since it reproduces correctly the asymmetry of
the measured curve.}
\end{figure}

Figure \ref{detail} (a) shows the diffraction curves of sample B-2
(20x etched\&annealed) measured in diffractions 002, 004 and 224
along with their fits assuming a (Ga,Mn)As layer with homogeneous
concentrations of Mn interstitial and substitutional atoms. The
quality of the fit is reasonably good, however a close inspection
in Fig. \ref{detail} (b) reveals that the fitted curve does not
reproduce well the asymmetry in the intensities of the $\pm$1st
thickness oscillations around the layer maximum. From this
asymmetry it follows that the (Ga,Mn)As layer in this sample is
vertically inhomogeneous. In order to obtain a better fit, two
sublayers with various densities of Mn interstitial atoms have to
be considered. The fitted experimental data using the model of one
layer and the model of two sublayers are shown on Fig.
\ref{detail}. Of course, the interstitial concentration profile is
a continuous function of $z$, however more than two layers in the
model would lead to an ambiguous solution of the fitting
procedure. This two layer approximation allows us to describe the
vertically inhomogeneous concentration as a step-like function of
the depth. The errors of the interstitial concentration in the
individual sublayers (mentioned in the Tab. \ref{tabulka}) is
comparable with the difference of their  values. We should
emphasize that these errors come from the uncertainty of the
\emph{average} interstitial density in the sample B-2 and the
difference in these interstitial concentrations is
$c_\mathrm{int}^{\mathrm{B-2,top}}-c_\mathrm{int}^{\mathrm{B-2,bottom}}=(0.15\pm
0.03)\%$, as follows from the fit. The presence of the
inhomogeneity in the interstitial concentration is demonstrated
unambiguously, since the uncertainty of the difference of the
concentrations is much smaller than the error of the total
interstitial density. The results of the fitting is depicted in
Fig. \ref{profil}.

The total thickness of the (Ga,Mn)As layer in sample B-2 was
determined from the HRXD analysis to $(87.0\pm 0.2)$ nm. From
the comparison of the thicknesses of samples B-1 and B-2 it
follows that the average decrease of the thickness due to the
rinsing in HCl for 30 seconds was 0.8 nm. This value corresponds
also to the difference between the thicknesses of the samples A
and B-1. Most probably, this is the thickness of a thin oxidized
layer at the surface corresponding to the Mn-rich layer after
annealing.

\section{\label{Simulation}Diffusion of Mn interstitials}

The results of the HRXRD measurements presented in the previous
section unambiguously demonstrate that the (Ga,Mn)As layer in
sample B-2 is slightly inhomogeneous, i.e., the Mn interstitials
exhibit a diffusion gradient. We performed a detailed simulation
of the out-diffusion of Mn interstitials during annealing, taking
into account their drift in an internal electric field and the
sample surface acting as a sink for the interstitials.

The interstitial Mn ions diffuse through the sample during the
annealing, the movement of the Mn ions is electrically compensated
by the simultaneous diffusion of holes. Simulating the Mn
diffusion, we restrict to one dimension denoting $z$ the
coordinate across the layer, because of the lateral homogeneity of
the layer. In our notation we define the origin of the coordinate
system ($z=0$) at the interface between the (Ga,Mn)As layer and
the buffer layer, the z-axis is pointing to the substrate. The
thickness of the layer is denoted by $H$, therefore, the
coordinate of the free layer surface is $z=-H$.

The diffusion is described by the following drift-diffusion
equations\cite{drift-diff-eq}
\begin{eqnarray}
\label{DDEn}
\frac{dn}{dt}=\frac{d}{dz} \left ( D_{n}\frac{dn}{dz} +
\mu_{n}n\frac{d\phi}{dz} \right )\equiv \frac{d}{dz}(-j_n), \\
\label{DDEp}
\frac{dp}{dt}=\frac{d}{dz} \left ( D_{p}\frac{dp}{dz} +
\mu_{p}p\frac{d\phi}{dz} \right ) \equiv \frac{d}{dz}(-j_p)
\end{eqnarray}
for the density of the Mn interstitials $n$ and the density of
the holes $p$. We define fluxes $j_{n(p)}$ of the interstitial
(hole) density according to the continuity equation. $D_{n}$ and
$D_{p}$ are the diffusion constants of interstitials and holes,
respectively, related to their mobilities $\mu_{n}$ and
$\mu_{p}$ by the Einstein relation\cite{Ashcroft}
\begin{equation}
D_{n,p} = \frac{\mu_{n,p} k_{\mathrm{B}}T}{q_{n,p}},
\end{equation}
where $T$ is the temperature, $k_{\mathrm{B}}$ is the Boltzmann
constant and $q$ is the electric charge of the corresponding
particle ($q=|e|$ for holes and $q=2|e|$ for interstitials) .

The electrostatic potential $\phi(z)$ fulfills the Poisson
equation
\begin{equation}
\label{potencial}\frac{d^2\phi}{dz^2}= \left ( c_\mathrm{sub}-p-2n \right )
\frac{e}{\epsilon},
\end{equation}
where $c_\mathrm{sub}$ is the concentration of the substitutional
Mn ions considered time-independent and homogenous within the
whole layer. The factor 2 on the right-hand side of the last
equation reflects the fact that the Mn interstitials are double
donors. On the right-hand side of the Eq. (\ref{potencial}) the
concentration of other charged defects, especially As antisite
defects, should be also included if they are present. In our case
however the density of the As antisite defects is very small
\cite{predchozi-clanek} -- definitely smaller than the estimated
error of the concentration of Mn acceptors (Mn atoms in
substitutional positions). The constants $e$ and $\epsilon$ are
the elementary charge and the permittivity of GaAs respectively.
The boundary condition for the electrostatic potential
\begin{equation}
\left.\frac{d\phi}{dz}\right|_{z\rightarrow\infty}=0
\end{equation}
follows from the zero electric field far below the (Ga,Mn)
As/GaAs interface.

The surface is passivated during the annealing due to the
diffusion of the oxygen from the air into the surface layer
\cite{PhysRevB.78.054403}, where Mn oxide is created. The
presence of the surface Mn oxide layer after the annealing and
its removal after the etching was indicated by the XANES
experiments, described later.

We model the process of the Mn oxidation as well as the diffusion
of the oxygen and the Mn atoms into the surface layer by a
phenomenological surface "container" at the sample free surface
$(z=-H)$ that traps the Mn interstitials coming from the volume of
the layer. The thickness of this container can be neglected since
the surface oxide layer is very thin (less than 3~nm), following
from other experiments \cite{PhysRevB.78.075319,
PhysRevB.78.054403}. The amount of the Mn ions is effectively
decreased just below the surface as some of them are transformed
to oxide losing their charge. We assume that the rate of change of
this amount is proportional to the local concentration of Mn
interstitials and to the remaining "free capacity" of the surface
container. Regarding these assumptions we define the flux of the
Mn interstitials into the surface container as
\begin{equation}
j_{n\mathrm{S}}(t)=S_0 \left ( 1 - \frac{N_\mathrm{S}(t)}{N_\mathrm{Smax}} \right )
n(t,z=-H),
\end{equation}
where the maximum capacity of the container is
$N_{\mathrm{Smax}}$, which corresponds to the possible maximum
number of the trapped Mn ions per unit area. Consequently, the
number of the trapped particles in the container in time $t$ is
\begin{equation}
N_\mathrm{S}(t)=\int_{-\infty}^{t}j_{n\mathrm{S}}(t) dt.
\end{equation}
The proportionality factor $S_{0}$ is connected to the rate of the
chemical reaction in the container (oxidation of Mn interstitials)
and it includes also the diffusion coefficient of the oxygen. The
value of this phenomenological parameter can hardly be estimated
from the theory. The behavior of the system was tested for various
values of $S_{0}$, i.e., from the limit of noninteracting surface
to the limit of the instantaneous oxidization of the Mn just below
the surface.

Since the Mn interstitial ion is a double donor, its oxidation at
the surface creates two holes and the flux of the interstitials
towards the surface is electrically compensated by the hole flux
into the (Ga,Mn)As layer
\begin{equation}
j_{p\mathrm{S}}=-2j_{n\mathrm{S}}.
\end{equation}

Realizing, firstly, that one boundary of our system is the surface
container, and secondly, there is no movement of the interstitials
or the holes far below the surface we can write the boundary
conditions for Eq. (\ref{DDEn}) and (\ref{DDEp}):
\begin{align}
\left.j_n\right|_{z=-H}=j_{nS}&,\quad \left.j_n\right|_{z\rightarrow\infty}=0,\\
\left.j_p\right|_{z=-H}=j_{pS}&,\quad \left.j_p\right|_{z\rightarrow\infty}=0.
\end{align}
In the calculation we consider zero fluxes of the interstitials
and holes in the depth comparable to the film thickness. Eqs.
(\ref{DDEn}) and (\ref{DDEp}) contain unknown parameters $D_{n}$,
$D_{p}$, $N_{\mathrm{Smax}}$ and $S_{0}$. We assume that these
quantities are independent on the local densities $n$ and $p$.

We assume that the post-growth annealing of the sample B-1 was
long enough (24~hours) to almost fill the container. Therefore,
the capacity of the surface container $N_{\mathrm{Smax}}$ is of
the same order of magnitude as the decrease in the total amount of
the interstitials during the first annealing, i.e., the difference
in the total amount of the interstitials in samples A and B-1.
This assumption is supported by the homogeneity of the layer in
sample B-1, following from the x-ray diffraction data. The absence
of a concentration gradient of the interstitials indicates that
the flux of the interstitials into the container
($j_{n\mathrm{S}}$) is smaller than their flux within the layer.
On the other hand, the flux into the container can be much larger
than the flux in the layer during the early period of the
annealing, when the container is empty. This leads to the vertical
inhomogeneity measurable after many etching and annealing steps
(sample B-2).

The mobility and the diffusion coefficient of the holes [$D_p\sim
10^{-3}~\mathrm{m^2/s}$ (Ref. \onlinecite{Sze})] at the annealing
temperature $160^\circ$C are assumed much larger than those of the
interstitials [$D_n\approx 1.4\times 10^{-21}~\mathrm{m^2/s}$
(extracted from the data in Ref.
\onlinecite{PhysRevLett.92.037201})]. Therefore, the movement of
the holes is rapid enough comparing to the slow interstitials so
that they are always in equilibrium for any configuration of the
interstitials. This assumption is valid if a difference of the
mobilities of the holes and the interstitials is at least several
orders of magnitudes, in this case the result of the simulation
does not depend on the value of $D_p$.

In our model, the rinsing in the HCl corresponds to the emptying
of the surface container. During the HCl dip the (Ga,Mn)As layer
is slightly thinned, this thickness variation is included in the
simulation of the cyclical cleaning\&annealing procedure. The
effect of this cyclical procedure transforming the sample B-1 to
B-2 was modelled by the simulation of the diffusion process.

\section{\label{Discussion}Simulation results and discussion}

In principle it is possible to optimize numerically the diffusion
parameters $D_{n}$, $N_{\mathrm{Smax}}$ and $S_{0}$ to fit the
experimental HRXRD data to the diffraction curves calculated from
the simulated concentration profiles. This approach is extremely
slow due to time consuming calculations of the diffusion, and
moreover, the sensitivity of the diffraction profiles to the
diffusion parameters is too small to obtain their precise values.
Therefore, we calculated the concentration profiles for many sets
of the diffusion parameters covering all possible combinations of
values within the limit of several orders of magnitude around the
expected values (estimated in Sec. \ref{Simulation} or previously
published\cite{PhysRevLett.92.037201}). We estimated the diffusion
parameters of the studied system by choosing the set of the
parameters bringing the best agreement of the simulated profiles
and the depth profiles determined by HRXRD. The concentration of
the interstitials resulted from the HRXRD measurements of the
sample A and B-1, resp. B-1 and B-2 determines the initial and
final states of the system for the simulation of the post-growth
and cyclical annealing, respectively.

Using this procedure, we found the following values $\mu_n \sim
2\times 10^{-18}~\mathrm{m^2\, V^{-1}\,s^{-1}}$, $D_n \sim 4\times
10^{-20}~\mathrm{m^2\,s^{-1}}$, and $S_0 \sim 3 \times
10^{-13}~\mathrm{m\,s^{-1}}$. The diffusion coefficient estimated
from our simulations is approximately 30 times larger than the
published value determined from electrical resistance
measurements.\cite{PhysRevLett.92.037201} If we used the published
value of the diffusion coefficient in our model, we were not able
to explain simultaneously both the observed inhomogeneity and the
decrease of the total amount of the interstitials. Unfortunately,
there is a lack of published values of the interstitial
diffusivity in (Ga,Mn)As for a detailed comparison. For instance,
the values of the diffusion coefficient of Mn interstitials
smaller by several orders of magnitude can be found in Ref.
\onlinecite{Adel2011JPCM} than the value extrapolated from the
results in Ref. \onlinecite{PhysRevLett.92.037201} for the same
annealing temperature($210^\circ$C), i.e., is higher than the
temperature used in our experiments (160$^\circ$C).

\begin{figure} 
\includegraphics[keepaspectratio,width=8cm,clip]{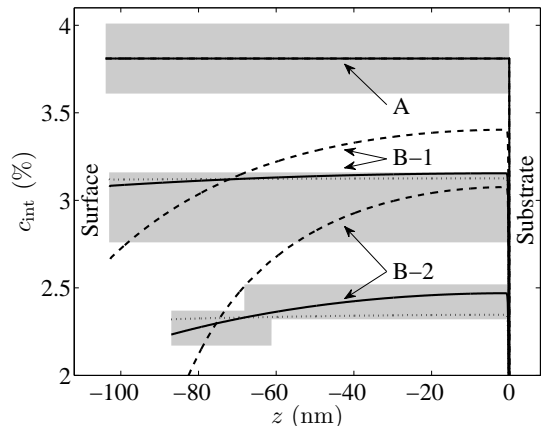}
\caption{\label{profil} Depth profiles of the concentration of the
Mn interstitials in sample A, B-1 and B-2. The concentration
profiles determined by HRXRD are represented by gray areas
indicating the uncertainty of the profiles. The profiles obtained
from the diffusion simulations for the interstitial diffusion
constant $D_n = 4\times 10^{-20}~\mathrm{m^2/s}$ are plotted by
solid lines; the concentration profiles simulated for 10 times
larger and 10 times smaller values of the Mn diffusion constant
are plotted by dotted and dashed lines, respectively. The initial
concentration profile for all simulations is given by the
concentration of Mn interstitials in sample A.}
\end{figure}

The comparison of the simulated concentration profiles and the
profiles determined from the HRXRD measurements are shown in Fig.
\ref{profil}. The uncertainty of the HRXRD results, which
evidently affects also the initial values for the simulation,
allows us to estimate the diffusivity only very roughly. For the
demonstration of the sensitivity of the profile on the diffusivity
value, we included also the simulation results for 10 times larger
and 10 times smaller diffusivities in Fig. \ref{profil}, the
latter value is close to that in
\onlinecite{PhysRevLett.92.037201}. A higher diffusivity leads to
a more homogenous concentration profile, whereas a smaller
diffusivity yields a highly nonhomogeneous profile.

As determined by HRXRD, the concentration of the interstitials
decreased by $\Delta
c_\mathrm{int}=c_\mathrm{int}^\mathrm{A}-c_\mathrm{int}^\mathrm{B-1}=(0.85\pm
0.12)$~\% during the post-growth annealing (A $\rightarrow$ B1).
From the decrease of the total amount of the interstitials it is
possible to evaluate the container filling
$N_{\mathrm{S}}^\mathrm{HRXRD} = (1.9 \pm 0.4)\times
10^{19}~\mathrm{particles/m^2}$. The optimized value of the
maximum container capacity following from the numerical simulation
of the diffusion is $N_\mathrm{Smax}\sim 3\times
10^{19}~\mathrm{particles/m^2}$. The estimated capacity of our
model container can be compared also to the published results,
\cite{PhysRevB.74.245205} where thin annealed (Ga,Mn)As layers
were studied by x-ray standing wave fluorescence. From this work
we used the value of the total Mn density in the 3~nm thick
surface layer after 4 hours of annealing at 200$^\circ$C and we
calculated the surface density of the interstitials in the
(phenomenological) surface container corresponding to this surface
layer. This density ($N_\mathrm{S}^\mathrm{XRSW}\approx 4\times
10^{19}~\mathrm{particles/m^2}$) is comparable in order of
magnitude with our value of then maximum container capacity.

Finally, from the diffraction data of sample B-3 it follows that
the layer was homogenized again during the final 80-hour annealing
(B-2$\rightarrow$B-3); the thickness of the crystalline layer is
smaller even without the etching procedure, most probably, due to
the strong oxidization of the surface during the long time delay
before the final annealing. Therefore, we are not able to
extrapolate the properties of the surface container to sample B-3
and this is the reason why the simulated and the measured profile
of the interstitial concentration in sample B-3 could not be
compared. The HRXRD measurement of sample B-3 indicates that the
inhomogeneity in the previous sample state (B-2) was not caused to
the formation of any immobile defects during the annealing.

\begin{figure} 
\includegraphics[keepaspectratio,width=8cm,clip]{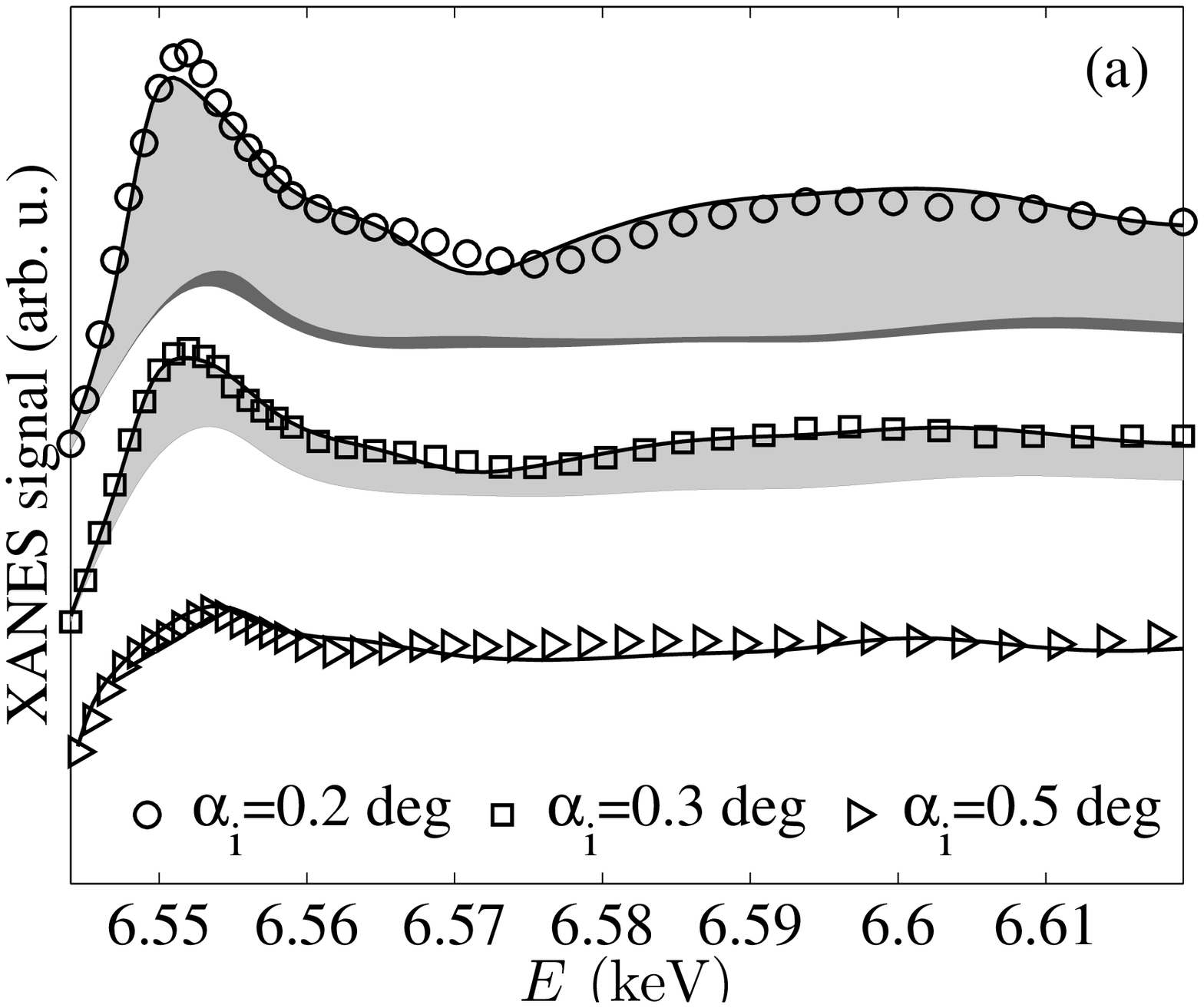}\\
\includegraphics[keepaspectratio,width=8cm,clip]{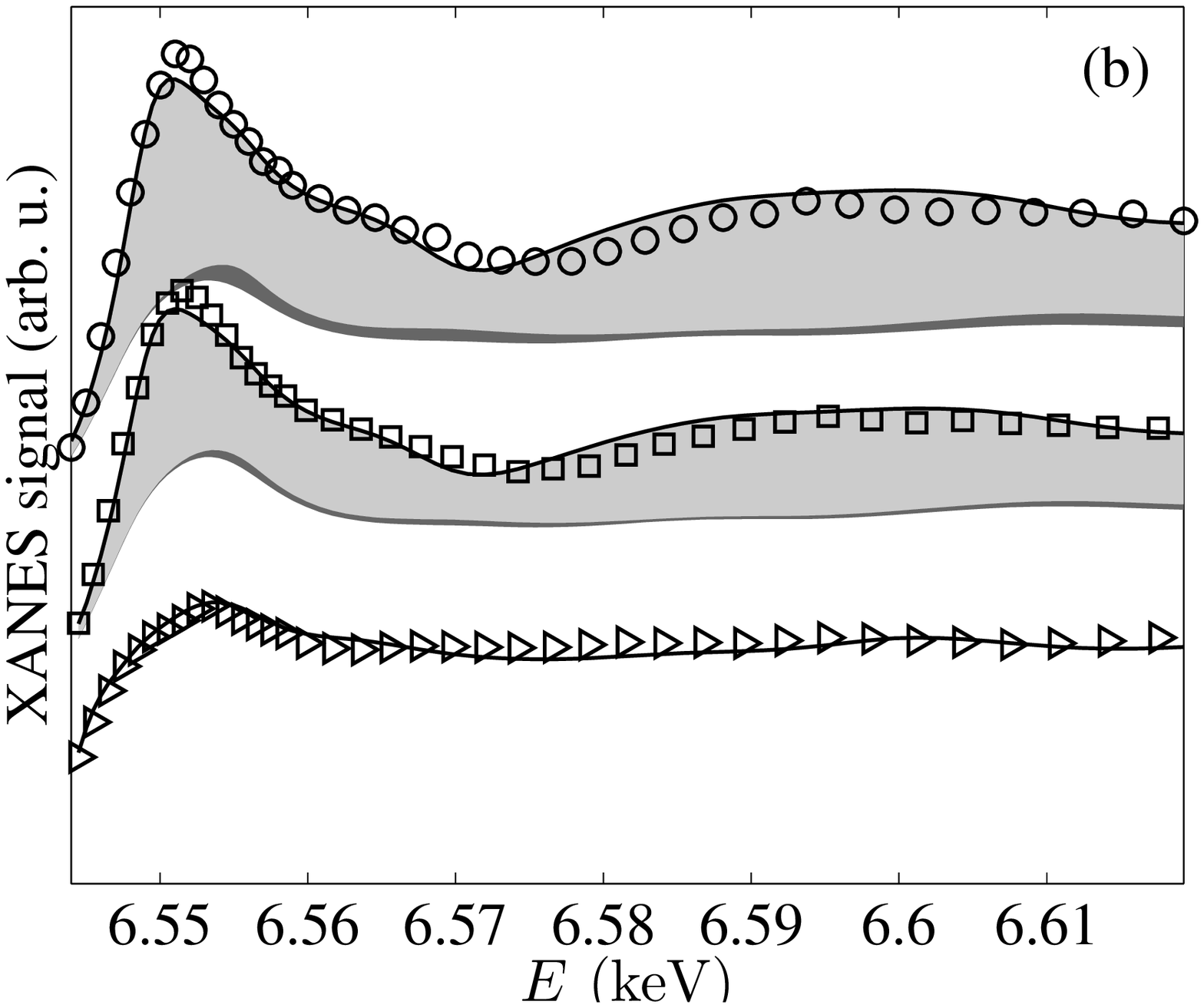}\\
\includegraphics[keepaspectratio,width=8cm,clip]{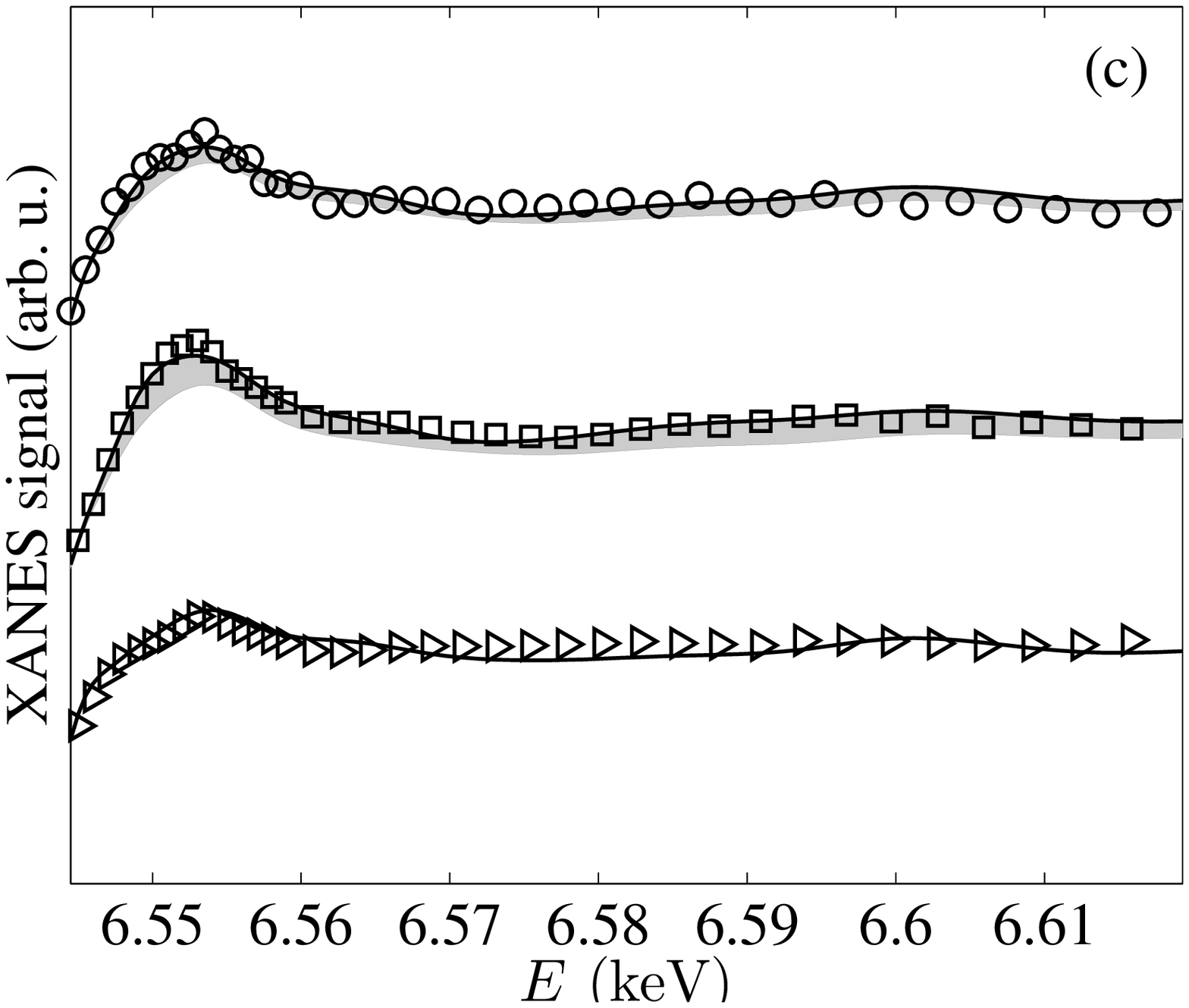}\\
\caption{\label{xanes-figs} XANES spectra obtained for as grown
sample (a), which was subsequently etched and annealed for one
hour (b) and finally etched again (c). The experimental data
(denoted by symbols) measured at various incidence angles
$\alpha_\mathrm{i}$ (i.e. for various penetration depths of the
primary beam) are shifted vertically for the clarity. The
calculated theoretical spectra fitting the experimental data are
plotted by solid lines, while the contributions of MnO and MnO$_2$
are emphasized by grey and dark grey areas, respectively.}
\end{figure}

In order to support our model of the surface container represented
by the surface Mn-oxide layer we performed a series of x-ray
absorption spectroscopy measurements around the MnK absorption
edge of the sample A in a reflection geometry with various
incidence angles of the primary beams, i.e., for various depths,
from which the XANES signal was collected. To observe the
influence of the etching and annealing on the surface, we did the
measurement for three modifications of sample A: firstly, the
original as grown sample A, secondly, the as grown sample A was
etched and annealed for one hour at 160$^\circ$C to create the
Mn-rich surface layer. Finally, the surface layer was removed by
another etching. The XANES spectra are displayed in Fig.
\ref{xanes-figs}.

The measured spectra were qualitatively compared with the results
of ab-initio XANES simulations using the FDMNES code.\cite{FDMNES}
We calculated the XANES spectra for Mn substitutional atoms, Mn
interstitial ions in two non-equivalent positions and for two Mn
oxide phases, namely cubic MnO (manganosite) and orthorombic
MnO$_2$ (groutelite). The measured spectra have been fitted to a
weighted average of the spectra simulated for various Mn
positions. From the data it follows indeed that the Mn-rich
layer\cite{PhysRevB.74.245205} at the surface consists of Mn
oxide. The presence of the Mn oxide can be clearly identified in
the as grown sample, this native oxide layer is very thin as the
significant XANES signal from the MnO (recognizable by the sharp
peak approx. at $E=6.55$ keV) was obtained \emph{only} for the
smallest angle of incidence [see Fig. \ref{xanes-figs}(a)]. After
the removal of the native oxide by etching and subsequent
annealing, the oxide layer was thicker than in the previous case,
since the intensity of the signal from the MnO is present also for
a larger penetration depth (i.e., for a larger angle of incidence)
(Fig. \ref{xanes-figs}(b)). Finally, from the fits of the
simulated spectra to experimental data [Fig. \ref{xanes-figs}(c)]
it follows that there is no Mn oxide at all after final etching of
the sample.

\begin{figure} 
\includegraphics[keepaspectratio,width=8cm,clip]{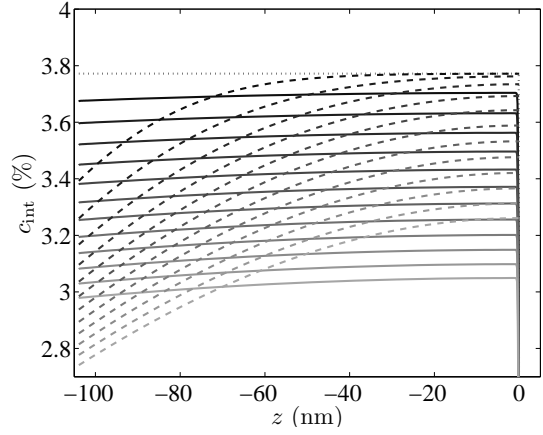}
\caption{\label{comparison} The depth profiles of the Mn
interstitial concentration following from the numerical
simulations for a post-growth annealing of sample A (as grown).
The solution of the drift-diffusion equations (solid lines) is
compared to the solution neglecting the drift and including only
diffusion (dashed lines). The time evolution of the depth profile
is represented by pitch of the line colors: from the darkest (the
earliest) to the lightest (final) colors with the time step of 2
hours. The dotted line shows the initial concentration profile of
uniformly distributed interstitials.}
\end{figure}

Concluding our observations from the numerical simulations we
formulate the interpretation of the annealing process in (Ga,Mn)
As layers, very often identified with out-diffusion of the
interstitials. The highly mobile holes reach very quickly an
equilibrium state, in which their diffusion flux is compensated by
the drift flux. This stable state is the result of two driving
forces: the compensation of the local charge unbalance and the
uniform distribution of the particles driven by the diffusion. If
the diffusion of the holes were neglected, the holes would
perfectly screen the charge of the substitutional Mn ions. In this
case of a perfect screening, the Mn interstitials would not feel
the electric field and their migration would be driven only by the
diffusion. The calculated profiles of the interstitial densities
including and neglecting the diffusion process of the holes are
compared in Fig. \ref{comparison}.

The excess holes near the surface caused by the passivation of the
interstitials, diffuse deeper to the layer leaving the space below
the surface negatively charged. On the other hand, the bottom part
of the layer starts charging positively as more holes arrive here
by the diffusion. This charge unbalance produces an electric field
acting on the interstitials. The resulting electrostatic force
acts on the interstitials in the same direction as on the holes,
since both particle types are positively charged. However, in
contrast to the holes, the surface layer acts as a sink for the
interstitials and consequently the interstitials diffuse to the
surface.

\begin{figure} 
\includegraphics[keepaspectratio,width=8cm,clip]{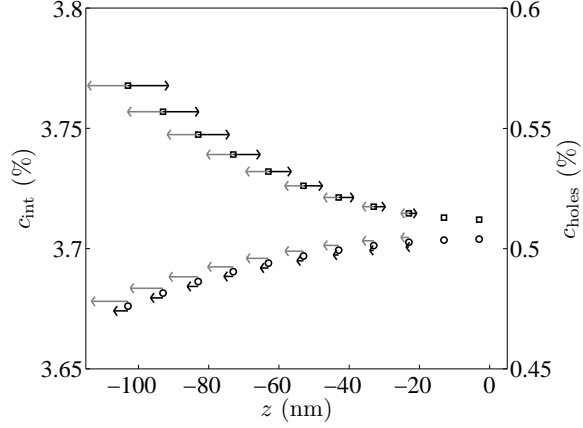}
\caption{\label{fluxes} The simulated depth profile of the
concentrations of the interstitials (circles) and the holes
(squares) after 2 hours of the post-growth annealing of sample A
(as-grown). The calculated fluxes caused by the diffusion (black)
and the drift (grey) are illustrated by the arrows indicating the
direction and the magnitude of the local flux. The arrows have
different scales for the holes and for the interstitials; for the
interstitials, the arrows for the diffusion flux are 10 times
magnified with respect to the arrows representing drift flux for
clarity.}
\end{figure}

Effects of the drift and the diffusion are superimposed creating
the resulting density profiles of the holes and interstitials
shown in Fig. \ref{fluxes}. From our simulations it follows that
the drift flux of the interstitials is larger than the diffusive
flux. As Fig. \ref{comparison} demonstrates, the diffusion of the
interstitials itself produce an inhomogeneous layer with a
\emph{large} concentration gradient. Such a large inhomogeneity
was not observed by HRXRD; the (Ga,Mn)As layers are usually
homogenous even for samples in which the annealing was stopped
before reaching the best possible improvement of their magnetic
properties. On the other hand, the model including both migration
processes of the interstitials and the holes produces a
\emph{small} gradient of the interstitial concentration and the
number of the interstitials in the layer after the annealing is
smaller than in case of the diffusion alone. Therefore, the main
process responsible for the interstitial migration is the drift
driven by the rapid diffusion of the holes.

\section{\label{Conclusion}Conclusion}

We studied the diffusion of Mn interstitials in (Ga,Mn)As during a
post growth annealing that is responsible for the improvement of
the magnetic properties of this material. By a combination of a
multiple short-time annealing with the removal of a thin surface
layer a nonzero gradient of the interstitial concentration was
created. We used high resolution x-ray diffraction for the
determination of the depth profile of the interstitial
concentration. We showed that the results of the diffusion
simulations are consistent with the x-ray diffraction
measurements. Basic parameters of the diffusion were estimated by
a comparison of the concentration profiles obtained by the
diffusion simulation with the profile determined from the x-ray
measurement. We were able to estimate a capacity and a trapping
rate of the surface layer absorbing out-diffused Mn interstitials
as well as the mobility of the interstitials. From our work it
follows that a main mechanism responsible for the removal of Mn
interstitals, e.g., improvement of the magnetic properties of
(Ga,Mn)As, is the drift of Mn interstitials to a surface driven by
a diffusion of holes. Consequently, the movement of the
interstitials during post-growth annealing is substantially
affected by the internal electrical field caused by an
inhomogeneous distribution of holes and positively charged
interstitials.

\begin{acknowledgments}
This work is a part of the research programme MSM 0021620834
financed by the Ministry of Education of the Czech Republic. The
work has been supported by the European Community's Seventh
Framework Programme NAMASTE under grant agreement number 214499.
The XANES experiment was carried out at synchrotron ANKA, Germany.
\end{acknowledgments}

\newpage
%

\newpage


\begin{table} 
\begin{ruledtabular}
\begin{tabular}{ccc}
Sample & thickness (nm) & $c_\mathrm{int}$ (\%)\\
\hline
A (as grown)& $103.9 \pm 0.2 $& $3.81 \pm 0.2$ \\
B-1 (annealed)& $103.0 \pm 0.2 $&$ 2.96 \pm 0.2$\\
\multicolumn{1}{c}{B-2(20x etch.\&ann.)} & \multicolumn{2}{c}{}\\
top sublayer& $22.3 \pm 3.5 $& $2.3 \pm 0.1 $\\
bottom sublayer& $64.7 \pm 3.5$ & $2.4 \pm 0.1$\\
B-3 (homogenized)& $84.4 \pm 0.1$ & $2.2 \pm 0.2$\\
\end{tabular}
\end{ruledtabular}
\caption{\label{tabulka} The thicknesses of the (Ga,Mn)As layers
and the concentrations $c_\mathrm{int}$ of the Mn interstitials in
measured samples obtained from x-ray diffraction. The vertically
inhomogeneous (Ga,Mn)As layer of sample B-2 is described by two
individual sublayers.}
\end{table}

\end{document}